\documentclass[10pt]{spie}
\usepackage[utf8]{inputenc}
\usepackage{authblk}
\usepackage[square,sort,comma,numbers]{natbib}
\usepackage{graphicx}
\usepackage{float}
\usepackage{gensymb}
\usepackage{url}
\usepackage{caption}
\usepackage{subcaption}
\usepackage{amssymb}
\usepackage{amsmath}
\usepackage{bm}
\usepackage{siunitx}
\usepackage[normalem]{ulem}
\usepackage{wrapfig}
\usepackage{color}
\usepackage{upgreek}
\usepackage{booktabs}
\usepackage{multirow}
\usepackage{pbox}

\usepackage{geometry}
\title{Early Commissioning Results of Slit Mask Integral Field Units on the Southern African Large Telescope}
\author[a,b,*]{Sabyasachi Chattopadhyay}
\author[a,c]{Matthew A. Bershady}

\affil[a]{South African Astronomical Observatory, 1 Observatory Rd, Observatory, Cape Town, 7925, South Africa}

\affil[b]{Centre for Space Research, North-West University, Potchefstroom 2520, South Africa}

\affil[c]{University of Wisconsin, Department of Astronomy, 475 North Charter Street, Madison, WI 53706, USA}

\begin{document}

\maketitle

\begin{abstract}

Three fibre feed integral field units (IFUs), called Slit Mask IFUs (SMI), are being developed in the SAAO fibre-lab for the Robert Stobie Spectrograph (RSS). The smaller, 200 micron fibre IFU (SMI-200) has 309 x 0.9 arcsec diameter spatial elements covering an elongated hexagonal footprint of 18 $\times$ 23 arcsec is now being commissioned. The larger, 300 (400) micron fibre IFU, SMI-300 (SMI-400), has 221 $\times$ 1.35 arcsec (178 $\times$ 1.8 arcsec) diameter spatial elements covering an on-sky area of 18 $\times$ 29 sq. arcsec (21 $\times$ 44 sq. arcsec). In all SMI units there are two groups of 13 fibres offset by roughly 50 arcsec on either side of the primary array to sample sky. SMI-200 provides a median spectral resolution of 2400 at H$\upalpha$ wavelengths in a low resolution mode simultaneously covering 370 to 740 nm. At a higher grating angles the SMI-200 delivers spectral resolution up to 10,000. A future red spectrograph arm for RSS will extend the simultaneous wavelength coverage up to 900 nm at a median resolution of 6000 for the same IFU. With this upcoming red arm and with the fibre-fed, near-infrared spectrograph NIRWALS on SALT, SMI-300 enables wavelength coverage from blue to NIR wavelengths at the same spatial resolution and footprint. The SMI units are inserted in the same fashion as the existing long-slit cassettes at the SALT focal plane. Prismatic fold mirrors direct the focal plane into the fibre IFU and then back into the RSS collimator after the fibres are routed 180 deg within the cassette and formatted into a pseudo-slit. Fold-prisms ensure that the spectrograph collimator continues to see the same focal plane. In this paper we report the laboratory characterization and on-sky commissioning-performance of the first Slit Mask IFU, SMI-200.

\end{abstract}


\section{Introduction}

The application of fibre-optic integral field units in astronomy is only of a few decades \cite{Allington-Smith_2002,bershady,giraffe}. IFUs in general have become a staple for most telescopes around the world \cite{kelz,virusp,virusw,SAMI,Sanchez,manga,vlt,kmos,muse, vimos}. Integral field spectroscopy (IFS) is an ideal mode for observing extended objects and relevant scientific questions such as the physics of interstellar medium, star forming regions, stellar populations and dynamics in and around galaxies \cite{cwi,manga,Sanchez,SAMI}. The Southern African Large Telescope (SALT) with its large aperture is ideal for observing extended sources, but a high \'etendue IFS \cite{chattopadhyay2022, wolf2022} is key to realizing the full potential of the telescope. One advantage of fibre-based IFS lies in the relative ease of retrofitting an existing wide-field spectrograph for new purpose. Such retrofits can avoid the relatively large expenditure of building an entirely new spectrograph. The SALT Slit Mask IFU (SMI) is an example of such retrofit system: These IFUs reformat a portion of the 8 arcmin field of RSS at SALT's prime focus for IFS. This is done with the relatively trivial swapping of mechanically-identical cassettes housing existing long-slit masks with the SMI cassettes. These cassettes are all loaded in a cartridge, selectable in real time to cater to specific programs dictaated by SALT's queue and weather conditions. While the retrofit itself is trivial at the telescope, the design and construction of the SMI cassettes and their optics is challenging, particularly given the very small volume inside the cassettes. We discuss the constraints the existing mechanism pose, and how we surmounted them to
ensure limited instrument development cost, rapid commissioning, and low probability of failure.

The SMI comes in three variants with 200, 300 and 400$\mu$m fibers. These are called SMI-200, SMI-300 and SMI-400 respectively; on SALT their fibers each subtend 0.9, 1.35 and 1.8 arcsec. In section 2, we  start with a brief review of the instrument, while section 3 and 4 are dedicated to describe in-lab and on-sky engineering commissioning respectively. We  conclude with lessons learned from the first unit (SMI-200) in the final section.

\section{Brief Instrument Description}

\begin{figure}[h]
    \centering
    \includegraphics[width=\linewidth]{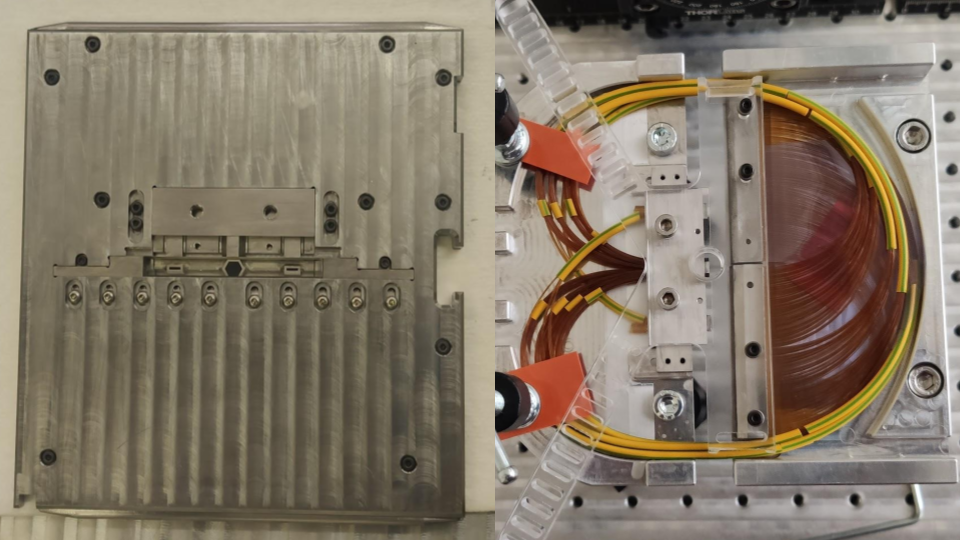}
    \caption{Fully assembled Slit Mask IFU SMI-200 with external features (left) and internal structure in a assembly jig (right). The total dimension of the SMI-200 is 134mm $\times$ 130mm $\times$ 8mm. Notice the upper loop (called right in the text) in the inner structure has noticeably smaller radius of curvature compared to the bottom loop (left in the text). The smaller radius of curvature led to stronger focal ratio degradation.}
    \label{fig:SlitMaskIFU}
\end{figure}

In this section we provide a brief overview of the Slit Mask Integral Field Units (SMI); a the detailed instrument description can be found in \cite{chattopadhyay2022}. The SMIs are a suite of interchangeable, retrofitted focal-plane masks for the Robert Stobie Spectrograph (RSS) on the Southern African Large Telescope (SALT). RSS sits at SALT's corrected prime focus, and accepts a host of variable width long-slits and multi-slit masks to select objects of interest. The first generation SMI with 200 $\upmu$m diameter core fibres (SMI-200) can be changed interchangeably with other masks and shares the cartridge case with the existing slit masks. A robotic elevator mechanism is used to fetch masks from their location in the case and insert them into the letter box shaped focal plane mechanism. Three wheels with slots on each side keep the masks at the focal plane within $\pm$100 $\mu$m accuracy along the optical axis during the observing track. To ensure compatibility, the SMI cassettes have outer dimensions of 134mm $\times$ 130mm $\times$ 8mm, identical to the long-slit cassettes. The thinness of the cassette forces the use of mirrors to fold the light beam from the telescope into the fibres, which then wrap 180$^\circ$ out of the focal plane within the volume mentioned above to another set of fold mirrors to redirect the light into the spectrograph. The fibre input is in the form of a 2D array (the object IFU), plus outlying fibres sampling sky. The fibre output is arranged in a linear pseudo-slit. The mechanical arrangement is shown in Figure \ref{fig:SlitMaskIFU}. The fibres are air-gapped with the prisms.

\begin{figure}[h]
    \centering
    \includegraphics[width=\linewidth]{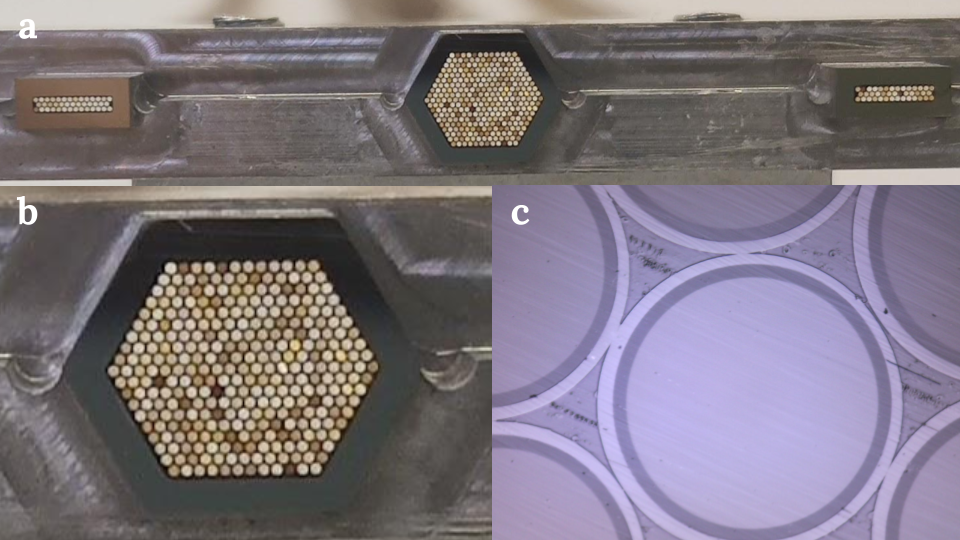}
    \caption{(a) Image of the fully polished SMI-200 (200$\upmu$\rm m diameter core fibre IFU)  object bundle (center) and the two sky arrays. (b) Zoomed in image of the object bundle. Note the excellent packing regularity. (c) Image of an individual fibre surrounded and in contact of other fibres demonstrating the tight packing of the array.}
    \label{fig:IFUface}
\end{figure}

SMI-200 contains 309 spatial elements (fibres) in the target, or object IFU, covering an extended hexagonal footprint of 18$\times$23 arcsec on sky, as shown in figure \ref{fig:IFUface}. To form this array, fibres are inserted into a metal ferrule with appropriate clearance to ensure regular packing with error within $\pm$7$\mu$m, similar in concept to what was done for the MaNGA IFUs \cite{drory15}. Each fibre collects light from a 0.9 arcsec diameter circular aperture. The hexagonal fibre array has a fill-factor of roughly 60\%. This fill-factor arises from the hexagonal packing and the 240 $\mu$m (1.06 arcsec) centre-to-centre spacing required by the fibre clad and buffer. To achieve 100\% spatial coverage requires dithered on-sky exposures in three defined positions. 

There are an additional 24 fibres in two offset locations to sample sky on either side of the hexagonal array. The sky fibres are arranged in two linear arrays, each spanning 15 arcsec, oriented on the long axis of the central hexagonal array. Each linear array is offset from the hexagonal array by roughly 70 arcsec (center to center), with edge-to-edge spacing of at least 50 arcsec. For objects that fill both the central hexagonal array and the sky arrays, it is necessary to have separate, offset exposures to sample sky. 

\begin{figure}[h]
    \centering
    \includegraphics[width=\linewidth]{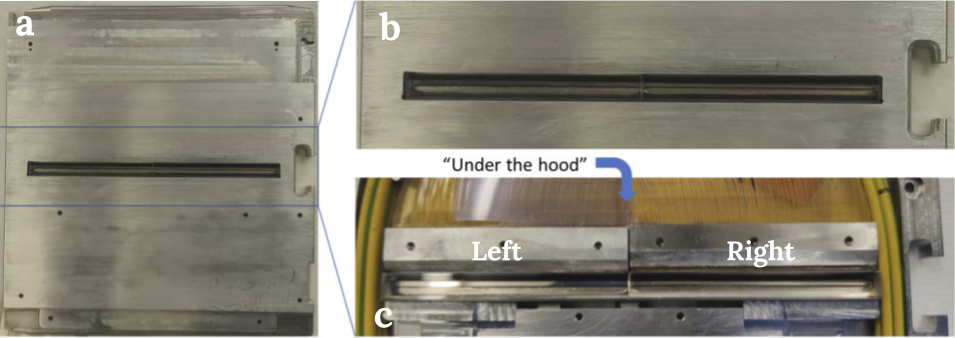}
    \caption{Completed article SMI-200 seen (a) looking “down” at the photon-exit side, i.e., what
    RSS sees. Visible is the 105mm length fibre pseudo-slit, expanded in (b), folded 90 deg by the prisms. (c) The inside of the cassette with the lid taken off (“under the hood”).}
    \label{fig:V-groove}
\end{figure}

The 333 fibres are arranged in a pseudo-slit made of two V-groove blocks that together spa roughly the 8 arcmin entrance aperture of the spectrograph (refer to figure \ref{fig:V-groove}. Given the fibre-to-fibre separation, each fibre produces a well-separated spectrum that is  extracted and wavelength calibrated as separate apertures. To optimise sky-subtraction, the sky fibres are carefully interleaved with the object fibres along the slit, and are positioned on sky to mimic the telecentric angle of the object fibres. 

Given the edge fibres face higher vignetting while the central fibres samples relatively brighter part of the galaxy we route the central objects fibres to the edge of the slit while the circumferential object fibres are mounted at the center of the slit. The object IFU is divided into two halves and they are routed to separate half slits via two narrow channels around the edges of the slit.

\section{Laboratory Characterization}

SMI-200 has had its optical performance characterized using a laboratory calibration engine. Measurements
include throughput and focal ratio degradation using an injection beam mimicking SALT's f/4.2 corrected prime focus. Despite the relatively fast injection speed, because the fibres inside SMI-200 case must bend tightly (refer to figure \ref{fig:SlitMaskIFU} to pass through narrow channels around the two edges of the pseudo slit, it is expected that the focal ratio degradation will be significant.

\begin{figure}[h]
    \centering
    \includegraphics[width=0.8\linewidth]{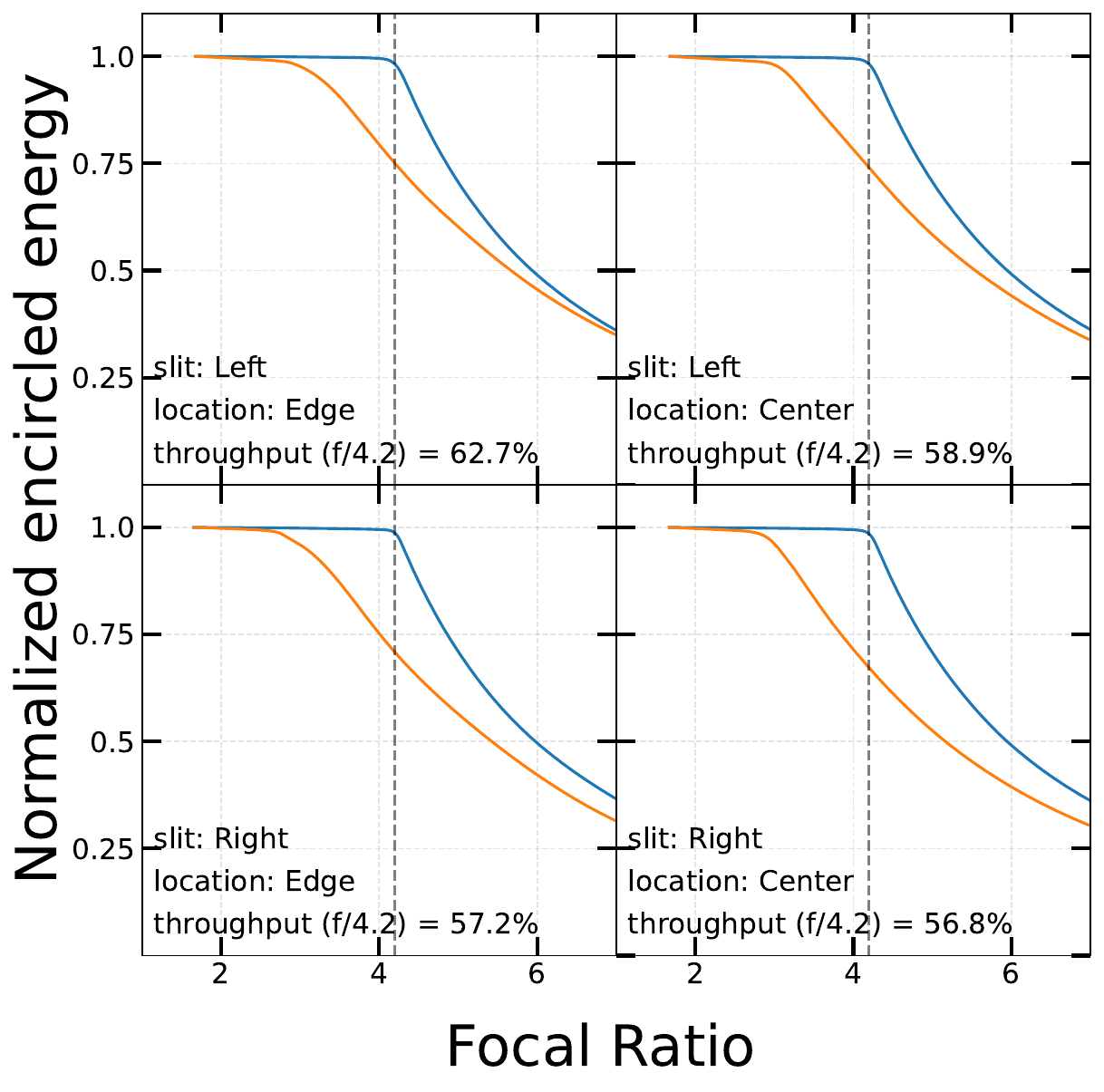}
    \caption{Focal ratio degradation performance of a sample of six fibres. Panels in the top and bottom rows are for fibres in the left and right half slits, respectively (bottom and top in Figure~\ref{fig:SlitMaskIFU}). Left panels show fibres at the center of the pseudo slit but edge of the IFU. Right panels show fibres at the edge of the pseudo slit but at the center of the IFU. Notice that left-half slit fibres have better encapsulated energy at f/4.2 compared to right-half slit fibres; the latter have smaller radius of curvature as shown in Figure~\ref{fig:SlitMaskIFU}.}
    \label{fig:frd}
\end{figure}

Our laboratory calibration engine is the SAAO-Wisconsin fibre Tester (SWiFT). A description and setup of SWiFT can be found \cite{chattopadhyay2022}. The spectrograph also accepts f/4.2 beam, hence the throughput was measured by comparing the flux between direct beam and fibre beam within the aperture set by f/4.2 focal ratio (relative throughput). The fibre redistributes a significant fraction ($\sim$25-35\%) of the input flux into higher angles at the fibre output end (refer to figure \ref{fig:frd}). Due  to the vignetting profile of the spectrograph, all of this light is lost and hence for all practical purposes, relative throughput is the important parameter. 

\begin{figure}[h]
    \centering
    \includegraphics[width=0.95\linewidth]{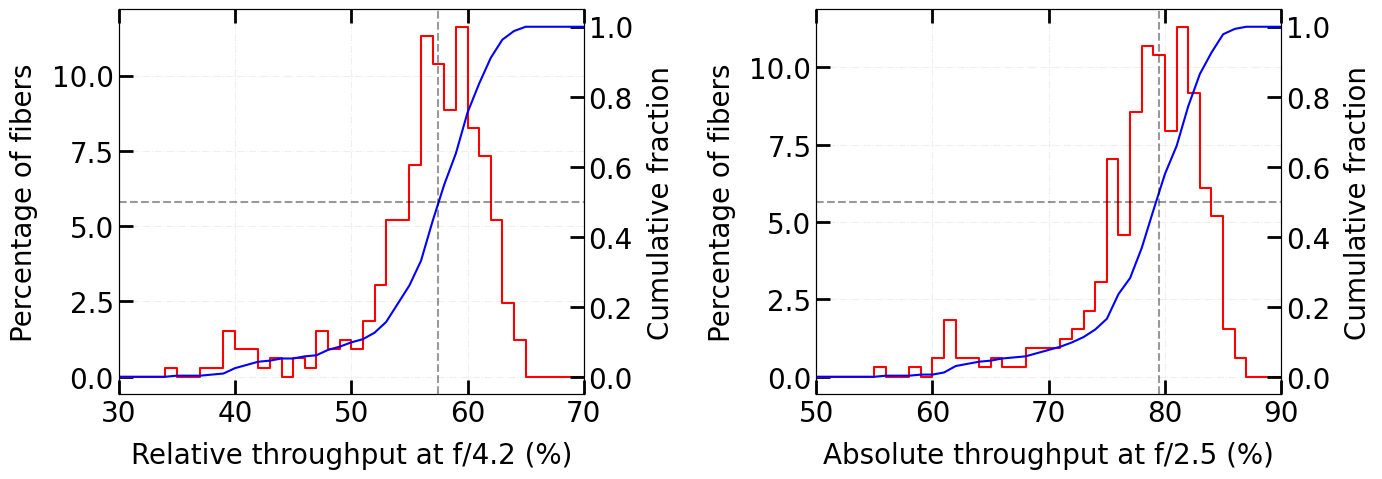}
    \caption{Histogram (left axis) and cumulative (right axis) numbers of relative (left panel) and absolute throughput (right panel) of SMI-200 fibres. The relative throughput is measured within an aperture set f/4.2 beam while the absolute throughput is measured within f/2.5 beam. The median relative throughput is found to be 58\% while median absolute throughput is 79.5\%.}
    \label{fig:tp_hist}
\end{figure}

Figure~\ref{fig:tp_hist} shows that the median \textit{relative} throughput of the fibres is 58\% at f/4.2, while the median \textit{absolute} throughput is 79.5\%. The absolute throughput is measured at an output beam speed of f/2.5, signifying the large focal ratio degradation introduced by the stressed bending situation inside the cassette. The median absolute throughput is also found to be very similar to reference single fibre throughput measurement (80\%) at the smallest bending radius of 50mm (\cite{chattopadhyay2022}). The 20\% drop in throughput can be expected as there is 7\% loss from Fresnel effects at both ends (total) of the fibre and 13\% from the two fold prisms (1\% loss at each of the four air-glass interfaces and 5\% loss at each of the two reflective surfaces).

\begin{figure}[h]
    \centering
    \includegraphics[width=0.95\linewidth]{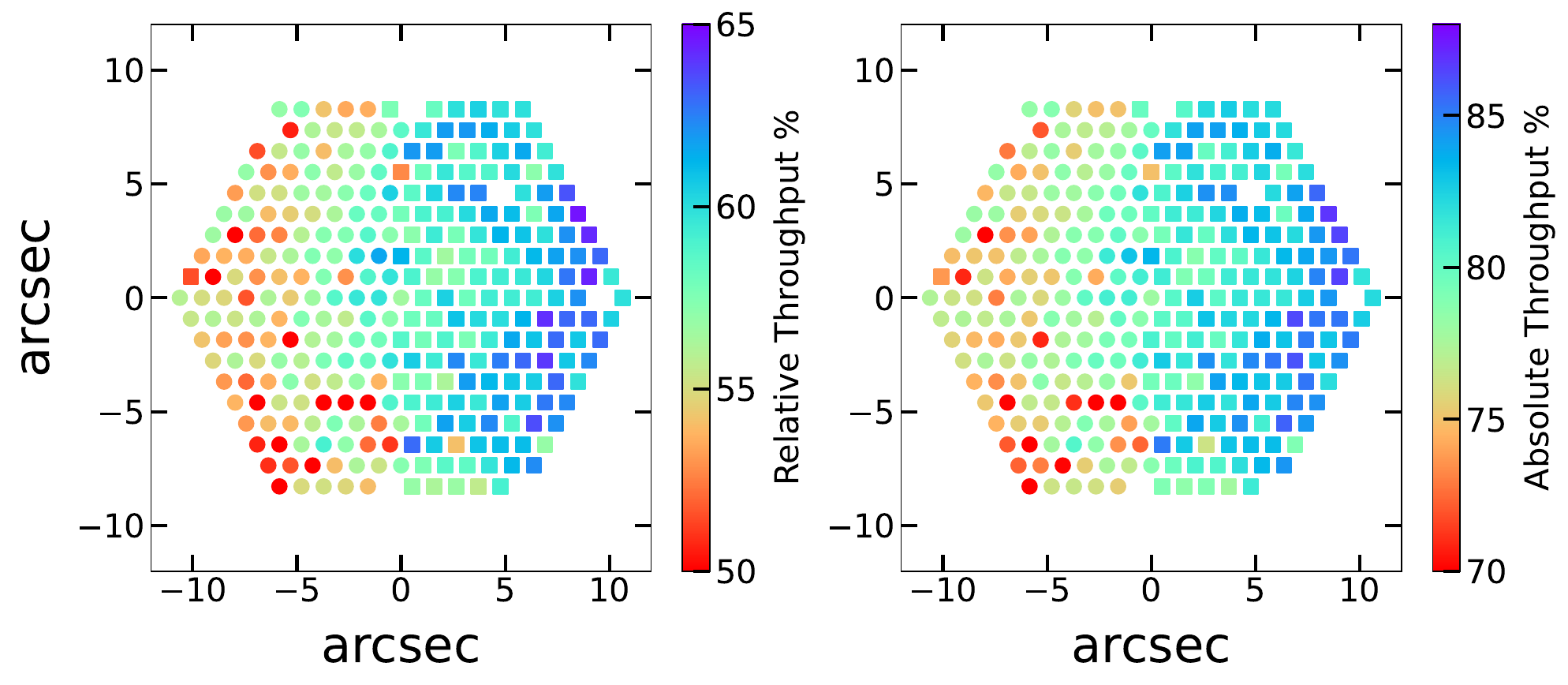}
    \caption{Spatial distribution of relative (left panel) and absolute (right panel) throughput. The blank fibres denote throughput less than 1\%. Note the fibres on the right side of the hexagonal array (fibres from the left slit, denoted by square symbol) have better throughput, on average, compared to the fibres on the left side (fibres from the right slit, denoted by circle symbol). The latter have a smaller radius of curvature in their routing to the fibre pseudo-slit, as shown in Figure \ref{fig:SlitMaskIFU}.}
    \label{fig:tp_map}
\end{figure}

We find the median relative throughput of the left half of the slit (60\%) is higher than the right half slit (56\%), as can be seen in Figure~\ref{fig:tp_map}. The lower median relative throughput directly correlates to smaller bending radius required to route the fibres to the right half of the slit, as seen in Figure~\ref{fig:SlitMaskIFU}. This can also be seen in the differences of the encircled energy profiles between left-half and right-half of the slit in Figure~\ref{fig:frd}. Top and bottom panels describe how fibres in left and right slit behave respectively. The two panels shows sample fibre behaviour for fibre located at the center, edge of the slit. While on average, the left slit fibres encompass 75\% of the total injected light at f/4.2, the right slit fibres only accepts 70\% of inject light at the same focal ratio. 

We also find that six fibres have extremely low throughput ($\leq$1\%) but are \textit{not} broken, based on direct physical inspection. Despite our attempts to route fibres in non-overlapping race-tracks with the help of thin-wall PVC tubing, these 6 fibres seem to have helically wrapped around other fibres over a small length. We suspect this has created, in these 6 cases, much smaller radii of curvature, leading to large throughput loss due to FRD.

\begin{figure}[h]
    \centering
    \includegraphics[width=0.95\linewidth]{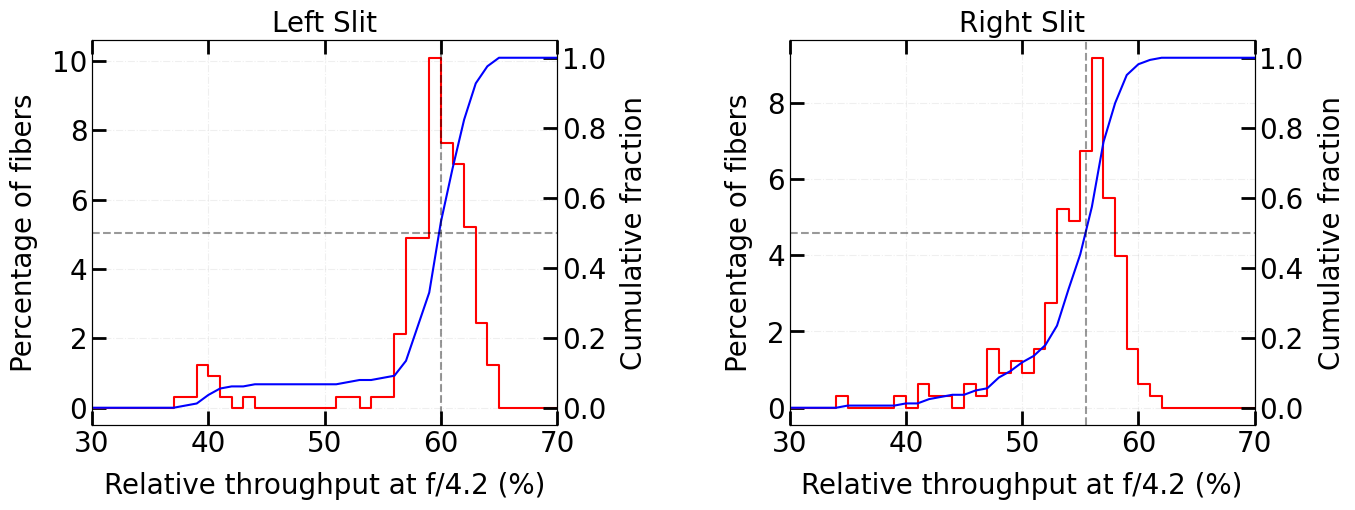}
    \caption{Differential (left axis) and cumulative (right axis) distributions of the relative fibre throughput. Fibres in in the left(right) side of the IFU hexagon are shown in the left(right) panel. The median relative throughput is 60\% for the left side while the same is 56\% for the right side which has a smaller bending radius as shown in figure \ref{fig:SlitMaskIFU}.}
    \label{fig:tp_hist_lr}
\end{figure}

\section{Preliminary on-sky commissioning result}

\subsection{Relative fibre throughput}
\begin{figure}[h]
    \centering
    \includegraphics[width=0.8\linewidth]{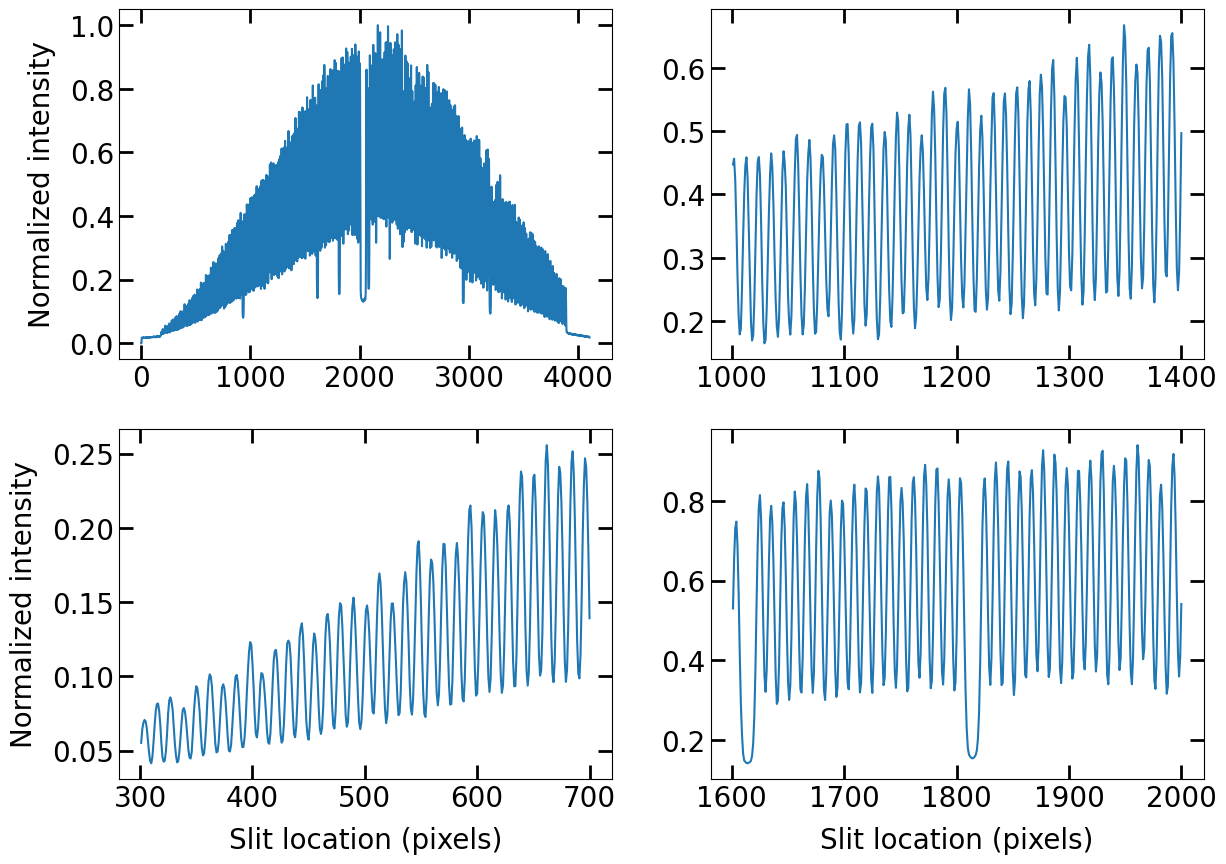}
    \caption{Cross-section (along slit axis) of continuum traces for SMI-200 fibres, as observed on-telescope with RSS. The top left panel shows the trace profiles near the wavelength midpoint for the full SMI-200 pseud-slit comprised of 327 fibres. The upper envelop represents the trace peaks; the lower envelop represents the inter-trace (inter-fibre) trough. The gap near pixel 2000 represents the gap between the two V-groove blocks. The strong trend in trace peak intensity with slit position, discussed in the text, is attributed to a combination of the spectrograph vignetting and inaccurate telecentric positioning of the fibres. The other panels provide expended views of different parts of the slit to demonstrate the fibre traces can be fitted with a trough to peak ratio of 0.5 or better. Gaps in the bottom right panels indicate presence of low throughput fibres.}
    \label{fig:slitilum}
\end{figure}

SMI-200 is currently undergoing commissioning at SALT. In this section we show some of the preliminary results from the first commissioning run. For these observations, RSS was configured to use the 900 l/mm volume-phase holographic grating (VPHg) with an incidence angle
of 15.125\degree\ grating angle, and a camera angle of 30.25\degree\ for a Littrow configuration. This captures a spectral range between 417 nm and 726 nm. We looked at fibre to fibre relative throughput using continuum (quartz-halogen) lamps from the SALT calibration system (CalSys). CalSys does not illuminate the SALT focal-plane uniformly, nor like the sky, but it is sufficiently uniform over the inner 50 arcsec (radius) to make an initial assessment of the relative fibre throughput. A cross-section of the continuum fibre traces are shown in Figure~\ref{fig:slitilum}. The top left panel shows traces along the full slit, while the other panels zoom in on different sections of the slit. Two primary attributes can be gleaned from this figure. First,
the fibre separation and camera optics lead to moderately well separated fibre traces with peak to trough variation yielding a contrast of 2 to 2.5. Second, there is a very strong 'vignetting' function as a function of slit position. This was not anticipated.

\begin{figure}[h]
    \centering
    \includegraphics[width=0.7\linewidth]{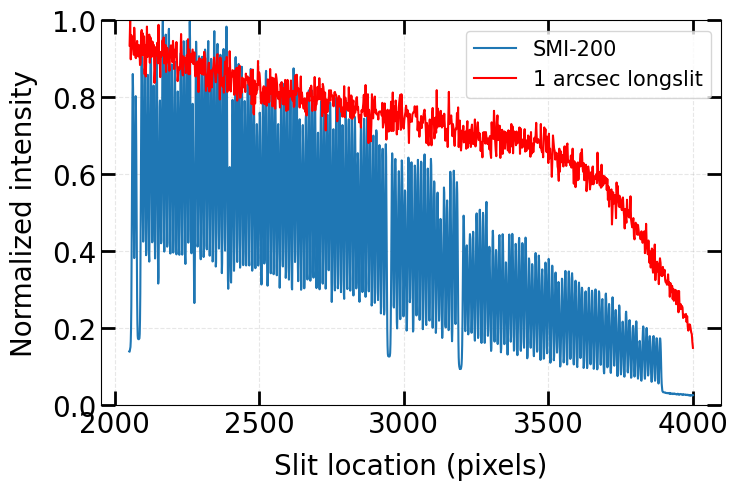}
    \caption{Relative fibre throughput of SMI-200 and 1 arcsec long-slit at 2000th pixel column along the spectral dimension for the left half slit. The effect of slit illumination variation is clearly comparable between SMI-200 and long slit relative illumination. The remainder of the effect (especially 2900 pixel onwards) can be attributed to the non-matching telecentricity.}
    \label{fig:slitcomp}
\end{figure}

To understand better the slit-profile for SMI-200, a direct comparison was made using the same spectrograph configuration and an imaging long-slit. The latter had a width of 1 arcsec. The slit function for one half of the long-slit and matching one-half of the SMI-200 pseudo-slit are shown in Figure~\ref{fig:slitcomp}. The strong variance from the center to the edge of the slit can be partly attributed to the slit illumination function while inaccuracy in telecentric angle match is found to cause the remainder of the effect. The relative throughput of SMI-200 from a flat field spectrum is found to be 30\% of 1 arcsec wide long slit. This is expected as the test fibre at fibre 163 have 50\% throughput while the fibre area is only 60\% of the long slit area.

\subsection{Telescope focus offset}

SALT long-slit cassettes have tilted and mirrored field stops to enable pick-off of the surrounding field for target acquisition via an Offner relay and a separate imaging camera. This option is unavailable with SMI-200 (or any of the SMI) because all of the available cassette volume is required for the fibre optics. We discuss the issue of acquisition below. The additional advantage of the slit-viewing mechanism is the ability to check and improve telescope focus. For the SMI we need to rely on  the long-slit masks for telescope focus, and determine if the fibre positions at the folded focal plane are par-focal with the long-slit apertures, as designed.

We tested the focus accuracy (par-forcality) by comparing the focus offset from a long slit. To do this we inserted a long slit mask at the focal plane and ensured the slit is focused on the telescope. We then replaced long slit with SMI-200 and took an exposure of a star at different telescope focus offsets in a range of $\pm$4mm. We found the number of fibres illuminated by the star is least when there is no focus offset. We performed this test at multiple locations of the telescope tracker with the same result. We conclude that SMI-200 internal focal plane coincide with the telescope focal plane within the tolerance of the letterbox mechanism.

\subsection{Spectrograph focus offset}

\begin{figure}[h]
    \centering
    \includegraphics[width=0.8\linewidth]{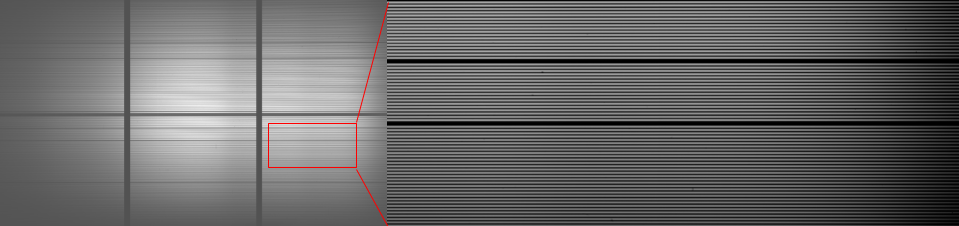}
    \caption{(Left) Flat field image with two QTH lamps (one for red wavelengths and the other for blue wavelengths) with 10 sec exposure time with a 20\% neutral density filter applied at 15.125\degree $~$grating angle and 30.25\degree $~$camera angle for the 900 l/mm VPHg. (Right) A zoom-in image of the red rectangle on the left demonstrating traces are well separated. Dark gaps indicate location of low throughput fibres.}
    \label{fig:flat}
\end{figure}

\begin{figure}[h]
    \centering
    \includegraphics[width=0.8\linewidth]{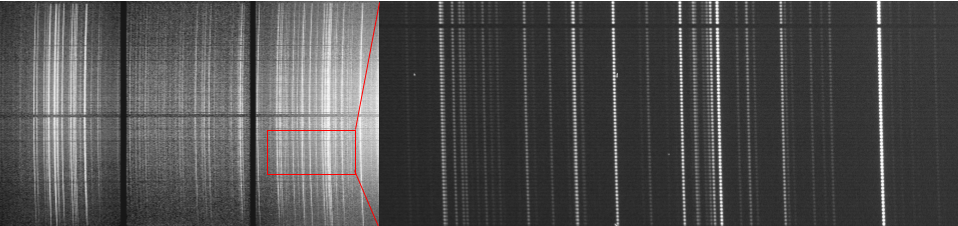}
    \caption{(Left) Arc lamp image with Xe lamp with 16 sec exposure time at 15.125\degree $~$grating angle and 30.25\degree $~$camera angle for the 900 l/mm VPHg. (Right) A zoom-in image of the red rectangle on the left panel. Dark traces indicate location of low throughput fibres.}
    \label{fig:arc}
\end{figure}

A similar issue concerns whether the fibre-output (pseudo-slit) is par-focal with the long-slit aperture. To to this, the spectrograph focus offset was computed against a long slit as well. We noted the nominal long slit focus value is determined by an analytical model based on temperature and location of the spectrograph on the tracker. We then replaced the long slit mask with SMI-200 and taken flat field exposures at multiple detector offsets from -100 to 600$\upmu$m at a step of 50$\upmu$m. At each step we measured the peak to trough ratio of fibre traces. We found that at 450$\upmu$m offset from the nominal long slit focus, the peak to trough ratio is maximized. Flat field spectra (refer to Figure \ref{fig:flat}) and arc spectra (refer to Figure \ref{fig:arc}) at 450$\upmu$m detector offset indicates that fibres are well focused on the detector. 

\subsection{Focal plane offset target acquisition}

\begin{figure}[h]
    \centering
    \includegraphics[width=0.7\linewidth]{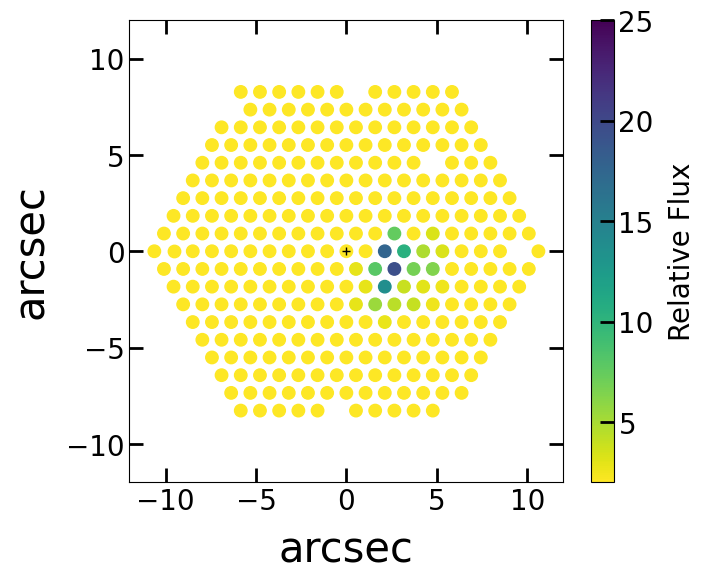}
    \caption{White light image of a star observed with SMI-200 after centering the star on the long slit. The black cross hair indicates the mechanical center of the IFU. The star is illuminating a set of fibres with a centroid having 2.7 arcsec x-offset and 0.95 arcsec y-offset. In this exposure the seeing was 2 arcsec, and the telescope appears to have drifted slightly over the track of 3 minutes.}
    \label{fig:white}
\end{figure}

The focal plane decenter of the IFU with respect to the long-slit cassettes was defined by observing a star. Again, the long slit mask is used as a reference. We first inserted a long-slit mask and centered the star on the long slit. We then replaced the long slit with SMI-200 in real time, and obtained an exposure. Fibre traces were fit, and continuum flux summed within the traces. Using the mapping of fibres from the IFU to the pseudo-slit, we created the continuum flux map shown in Figure~\ref{fig:white}. This image is used to centroid the star location within the IFU. We consistently measured 2.7 arcsec x-offset and 0.95 arcsec y-offset as can be seen in figure \ref{fig:white}. This offset is unaffected by SALT's tracker location and (re)insertion of SMI=200 into the focal plane via the letter box mechanism. 

\section{Summary}

We present the results from laboratory characterization and then early engineering commissioning of SMI-200, the 200$\mu$m fibre IFU for SALT's Robert Stobie Spectrograph. Lab measurements established the mapping between fibre locations in the IFU and the fibre pseudo-slit, and revealed that the instrument has 327 active fibres with median relative throughput of 58\% within an aperture of f/4.2. The remaining 6 fibres are not broken but demonstrate poor throughput due to sharp bending radius introduced by the compact design. The median absolute throughput at f/2.5 output focal ratio is found to be 80\%, which is expected assuming minimal transmission loss with the short length of fibre, but with 7\% and 13\% Fresnel losses at fibre and prism surfaces respectively. Roughly half of the Fresnel losses could be eliminated by optically bonding the fibers to the prisms with uncoated surfaces. The left side of the slit demonstrate better median relative throughput (56\% vs 60\%) compared to the right side which has a smaller fibre bending radius.

During on-sky commissioning, we found there is a strong trend in intensity along the length of the pseudo-slit, peaking in the center. Fibre traces can be extracted with peak to trough ratio of 2-2.5 across all wavelengths and pseudo-slit locations. The slit intensity trend is not due to telescope illumination effects, as is the case for the long-slit intensity profile. Instead, the pseudo-slit intensity profile is due to a mismatch in the telecentricity of the fibre pseudo-slit. This will be corrected in the second-generation SMI. The SMI-200 absolute on-sky throughput at fibre aperture 163 (near the center of the slit) is found to be $\sim$50\% of the long-slit throughput, accounting for
differences in entrance solid angles. This is consistent with light losses due to FRD and surface reflections from prisms and fibres.

We find that the SMI-200 can be inserted into the telescope and RSS focal plane without any telescope refocus required relative to long-slit cassettes. However, a 450$\upmu$m offset in the spectrogrpah focus is required relative to the long-slits. The spectrograph focus offset for long-slit cassettes is dependent on the temperature and track location but the offset between SMI-200 and any long-slit cassette remains fixed. This is consistent with the SMI-200 moving the input focal plane of the spectrograph by 450$\upmu$m. Continuum images of a star observed with the SMI-200 indicate that there is a 2.86 arcsecs radial offset between SALT focal and SMI-200 field centers.

\color{black}
\acknowledgements Support for this research has been provided by the South African National Research Foundation SARChI-114555.

\bibliography{report} 
\bibliographystyle{spiejour}

\end{document}